\documentclass[twocolumn,showkeywords,prb]{revtex4}
\usepackage{graphicx}
\usepackage{dcolumn}
\usepackage{amsmath}
\usepackage{latexsym}

\begin{document}
\title
{Signatures of noncommutativity in bar detectors of gravitational waves}
\author{
{\bf {\normalsize Sunandan Gangopadhyay}$^{a}$\thanks{sunandan.gangopadhyay@gmail.com,
sunandan.gangopadhyay@bose.res.in}},
{\bf {\normalsize Sukanta Bhattacharyya}$^{b}$,\thanks{sukanta706@gmail.com}}
{\bf {\normalsize Anirban Saha}$^{b}$\thanks{anirban@associates.iucaa.in}}\\
$^{a}$ {\normalsize  Department of Theoretical Sciences,}\\{\normalsize S.N. Bose National Centre for Basic Sciences,}\\{\normalsize JD Block, 
Sector III, Salt Lake, Kolkata 700106, India}\\[0.1cm]
$^{b}$ {\normalsize Department of Physics, West Bengal State University, Barasat, Kolkata 700126, India}\\}

\begin{abstract}
\noindent The comparison between the noncommutative length scale $\sqrt{\theta}$ and the length variation $\delta L=h L$, detected in the GW detectors indicate that there is a strong possibility to detect the noncommutative structure of space in the GW detector set up. We therefore explore how the response of a bar detector gets affected due to the presence of noncommutative structure of space keeping terms upto second order in the gravitational wave perturbation ($h$) in the Hamiltonian. Interestingly, the second order term in $h$ shows a transition between the ground state and one of the perturbed second excited states that was absent when the calculation was restricted only to first order in $h$. 
\end{abstract}

\maketitle

\section{Introduction}
\noindent The existence of an uncertainty in the spatial coordinates \cite{Dop, Alu} due to a sharp localization of events in space at the the quantum level is strongly supported by various gedanken experiments. The standard way to impose this uncertainty is to postulate the noncommutative (NC) Heisenberg algebra \cite{doug}
\begin{eqnarray}
	\left[{\hat x}_{i}, {\hat x}_{j}\right] =i \theta \epsilon_{ij}, \left[{\hat x}_{i}, {\hat p}_{j}\right] =i \hbar \delta_{ij},~\left[{\hat p}_{i}, {\hat p}_{j}\right] = 0 ; i, j= 1,2
\label{1}
\end{eqnarray}
where $\theta$ is the NC parameter and $\epsilon_{ij}$ is an anti-symmetric tensor with $\epsilon_{12}=1$. With this granular structure of space, NC quantum field theory (NCQFT) \cite{doug,SW}, NC quantum mechanics (NCQM) \cite{nair}-\cite{sb2} and NC gravity \cite{grav}-\cite{sgrb} have been constructed. A part of the endeavor has also been spent in finding the order of magnitude of the NC parameter and exploring its connection with observations \cite{mpr}-\cite{stern}. Studies in NCQM suggest that the NC parameter associated with different particles may not be the same \cite{pmho, vassilavich} and this upperbound in length scale could be as high as $\sqrt{\theta}\sim 10^{-20}m-10^{-17}m$ \cite{stern}.

\noindent Length scales of this order appear in a completely different context. The direct detection of the gravitational waves (GWs) \cite{ligo,ligo2} by the advanced LIGO detector \cite{ligoo}  has opened a new window to observe variations in length-scales that has never been probed before. Among the currently operating GW detectors \cite{bar_2}-\cite{bar_detectors_5} (LIGO \cite{abramovici}, VIRGO \cite{caron}, GEO \cite{luck} and TAMA \cite{ando}) where interferometric techniques are being used, one can detect a length-variation of the order of $\frac{\delta L}{L} \sim 10^{-23} $. Interestingly, the upper bound on the spatial NC parameter also has a length scale of this order ($\sqrt{\theta}\approx 10^{-20}$ m).
\noindent This motivates us to anticipate that a good possibility of detecting the NC structure of space would be in the GW detection experiments. It turns out that the response of a resonant bar detector to GW can be quantum mechanically described as GW-harmonic oscillator (GW-HO) \cite{Magg} interaction.

\noindent To predict the possible presence of NC structure of space in GW detection scenarios we had studied various aspects of the GW-HO interaction in NCQM framework in \cite{ncgw1}-\cite{sb}. In these studies we have worked out the response of the system which indicates that the spatial noncommutativity introduces a characteristic shift in the frequency where the HO will resonate with the GW. In \cite{sb} we worked out the complete perturbative calculation including both the time independent and time dependent interacting Hamiltonian to obtain transition probabilities among the shifted energy levels for a generic GW wave-form upto first order in $h$. Here, we investigate how the transitions get affected due to second order terms in $h$. Interestingly, we observe that there is no contribution from the $\mathcal O(h^2)$ terms in the transition probabilities calculated earlier in \cite{sb}. It only gives rise to another transition between the ground state to one of the perturbed second exited states of the HO. This transition was absent at first order in $h$ in our earlier study.
\section{Interaction between NC HO-GW }
\noindent In a resonant bar detector the phonon mode excitations \cite{Magg} interacting with the incoming GW, behave like a quantum mechanical forced HO. We therefore set out to analyse a quantum mechanical forced HO to explore the theory of interaction of GW with the resonant bar detector. Here we consider the effect of the GW to be constrained in the $2$-$D$ plane (taken in the $x-y$ plane) perpendicular to the direction of propagation of the GW. 

\noindent To construct the theoretical framework of our model, we write the geodesic deviation equation for a $2$-$D$ HO of mass $m$ and intrinsic frequency $\varpi$ in a proper detector frame as
\begin{equation}
m \ddot{{x}} ^{j}= - m{R^j}_{0k0} {x}^{k} - m \varpi^{2} x^{j}
\label{e5}
\end{equation}
where ${R^j}_{0k0} = - \frac{d \Gamma^j_{0k}}{d t}  = -\ddot{h}_{jk}/2  $,  ${R^j}_{0k0}$ denotes the components of the curvature tensor in terms of the metric perturbation $h_{\mu \nu} $ as $g_{\mu\nu} = \eta_{\mu\nu} + h_{\mu\nu}$; 
$|h_{\mu\nu}|<<1$ on the flat Minkowski background $\eta_{\mu\nu}$.

It is to be noted that here the transverse-traceless (TT) gauge has been used to remove the unphysical degrees of freedom. The TT gauge choice gives only two physical degrees of freedom, namely, the $\times$ and $+$ polarizations of the GW.
A convenient form of $h_{jk}$ showing these polarizations reads
\begin{equation}
	h_{jk} \left(t\right) = 2f \left(\varepsilon_{\times}\sigma^1_{jk} + \varepsilon_{+}\sigma^3_{jk}\right)
	\label{e13}
\end{equation}
where $2f$ is the amplitude of the GW, $\sigma^1$ and $\sigma^3$ are the Pauli spin matrices, and the two possible polarization states of the GW, $\left( \varepsilon_{\times}, \varepsilon_{+} \right)$ are restricted to follow the condition $\varepsilon_{\times}^2+\varepsilon_{+}^2 = 1$ for all $t$. 

\noindent The Lagrangian which leads to the geodesic deviation equation (\ref{e5}) is given by
\begin{equation}
{\cal L} = \frac{1}{2} m\dot {x_{j}}^2 - m{\Gamma^j}_{0k}
\dot {x}_{j} {x}^{k}  - \frac{1}{2} m \varpi^{2} x_{j}^2 ~.
\label{e8}
\end{equation}
\noindent This immediately leads to the Hamiltonian 
\begin{equation}
	{H} = \frac{1}{2m}\left({p}_{j} + m \Gamma^j_{0k} {x}^{k}\right)^2 + \frac{1}{2} m \varpi^{2} x_{j}^2 ~~.
	\label{e9}
\end{equation}
In this paper we set out to find the signature of the spatial noncommutativity considering the effect of second order terms in the perturbation $h_{ij}$ appearing in our model. To carry this out, first we elevate the phase-space variables $\left( x^{j}, p_{j} \right)$ to operators $\left( {\hat x}^{j}, {\hat p}_{j} \right)$ and impose the NC Heisenberg algebra. The NC operators are connected to the operators $X_{i}$ and $P_{i}$ satisfying the standard ($\theta=0$) Heisenberg algebra through 
\begin{eqnarray}
	{\hat x}_{i} = X_{i} - \frac{1}{2 \hbar}
	\theta \epsilon_{ij} P_{j}~, \quad {\hat p}_{i} = P_{i}~.
	\label{e9b}
\end{eqnarray}
Using this connection, the Hamiltonian in eq.(\ref{e9}) can be written in terms of the commutative variables upto second order in $\Gamma$ as
\begin{eqnarray}
	{\hat H} &=& \left(\frac{ P_{j}{}^{2}}{2m} + \frac{1}{2} m \varpi^{2} X_{j}{}^{2} \right)+ \Gamma^j_{0k} X_{j} P_{k} -\frac{m \varpi^{2}}{2 \hbar} \theta \epsilon_{jm} X^{j} P_{m} 
	\nonumber \\ &&
	-\frac{\theta }{2 \hbar} \epsilon_{jm} P_{m} P_{k}  \Gamma^j_{0k} +\frac{m}{2} \Gamma^j_{0k} \Gamma^j_{0l} X^{k} X^{l} +  \mathcal O(\theta \Gamma^2)\nonumber\\&& + \mathcal O(\theta^2) ~.	\label{e12}
\end{eqnarray}
\noindent The above Hamiltonian  in terms of the raising and lowering operator can be  recast as
\begin{eqnarray}
	{\hat H} &=&\hbar \varpi (a_j^\dagger a_j+1) - \frac{i\hbar}{4} \dot h_{jk}(t) \left(a_j a_k - a_j^\dagger a_k^\dagger\right) \nonumber\\ &&
	+ \frac{m \varpi \theta}{8} \epsilon_{jm} {\dot h}_{jk}(t)  \left(a_{m}a_{k}  - a_{m}a_{k}^\dagger + C.C \right) \nonumber\\ &&
	+ \frac{\hbar}{4 \varpi}\dot h_{jk}(t)  \dot h_{jl}(t) \left(a_{k}a_{l}+a_{k}a_{l}^\dagger+a_{k}^\dagger a_{l}+a_{k}^\dagger a_{l}^\dagger \right)
	\nonumber\\ &&
	-\frac{i}{2} m \varpi^2 \theta \epsilon_{jk} a_j^\dagger a_k
	\label{e16}
\end{eqnarray}
where C.C. means complex conjugate. The raising and lowering operators in terms of the oscillator frequency $\varpi$ is given by 
\begin{eqnarray}
X_j = \sqrt{ \frac{\hbar}{2m \varpi}}\left(a_j+a_j^\dagger\right);
P_j = \sqrt{ \frac{\hbar m\varpi}{2i}}
\left(a_j-a_j^\dagger\right).
\label{e15}
\end{eqnarray} 
We now write the Hamiltonian (\ref{e16}) into three parts as
 \begin{eqnarray}
 \hat{H}&=&\hat{H_0}+\hat{H_1}(t)+\hat{H_2}\nonumber\\
 \hat{H_0} &=& \hbar \varpi (a_j^\dagger a_j+1) \nonumber\\
 \hat{H_1}(t) &=&  - \frac{i\hbar}{4} \dot h_{jk}(t) \left(a_j a_k - a_j^\dagger a_k^\dagger\right) \nonumber\\&&
 + \frac{\Lambda}{4} \hbar \epsilon_{jm} {\dot h}_{jk}(t)  \left(a_{m}a_{k}  - a_{m}a_{k}^\dagger + C.C. \right) \nonumber\\&&
 + \frac{\hbar}{4 \varpi}\dot h_{jk}(t)  \dot h_{jl}(t) \left(a_{k}a_{l}+a_{k}a_{l}^\dagger+a_{k}^\dagger a_{l}+a_{k}^\dagger a_{l}^\dagger \right) \nonumber\\
 \hat{H_2} &=&  - i \Lambda_{\theta} \hbar \epsilon_{jk} a_j^\dagger a_k  
 \label{hbp}
 \end{eqnarray}
where $\Lambda=\frac{m \varpi \theta}{2 \hbar}$, 
$\Lambda_{\theta} =  \frac{m \varpi^2 \theta}{2 \hbar}$.



\noindent The perturbed energy states incorporating the effect of the time independent perturbation $\hat{H_2}$ was obtained in \cite{sb}. These read  
\begin{eqnarray}
	\psi_2^{(0)} &=& (| 2,0\rangle + | 0,2\rangle) \nonumber\\
	\psi_2^{(1)} &=& (| 2,0\rangle - | 0,2\rangle + i \sqrt{2} | 1,1\rangle ) \nonumber\\
	\psi_2^{(2)} &=& (| 2,0\rangle - | 0,2\rangle - i \sqrt{2} | 1,1\rangle )
	\label{f1}
\end{eqnarray}
with the the corresponding energy eigenvalues  
\begin{eqnarray}
	E_2^{(0)} &=& 3 \hbar \varpi \nonumber\\
	E_2^{(1)} &=& 3 \hbar \varpi (1 + \frac{2}{3} \Lambda) \nonumber\\
	E_2^{(2)} &=& 3 \hbar \varpi (1 - \frac{2}{3} \Lambda)~.
	\label{ev}
\end{eqnarray}
\noindent Now we proceed to compute the transition probabilities between the ground state and the perturbed non-degenerate second excited states of the $2-D$ harmonic oscillator using time dependent perturbation theory incorporating the second order correction in $h$. The probability amplitude of transition from an initial state $|i\rangle$ to a final state $|f \rangle$, ($i\neq f$), due to a perturbation $\hat{V}(t)$, to the lowest order of approximation in time dependent perturbation theory is given by \cite{kurt}
\begin{eqnarray}
	C_{i \rightarrow f}(t\rightarrow \infty)  =  -\frac{i}{\hbar} \int_{-\infty}^{t\rightarrow +\infty} dt' e^{\frac{i}{\hbar}(E_f -E_i)t'} \nonumber\\
	\times \langle \Phi_f | \hat{V}(t')|\Phi_i \rangle~.
	\label{probamp}
\end{eqnarray}
\noindent Using the above result, we observe that the probability of transition  survives only between the ground state $|0,0\rangle$ and the perturbed second excited states given by  eq.$(\ref{f1})$
where $\hat V(t')$ is given by
\begin{eqnarray}
	\hat V(t') &=& - \frac{i\hbar}{4} \dot h_{jk} (t')\left(a_j a_k - a_j^\dagger a_k^\dagger\right) + \frac{\Lambda}{4}  \hbar \epsilon_{jm} \dot h_{jk} (t')  \left( a_{m}a_{k} \right. \nonumber\\&& \left. - a_{m}a_{k}^\dagger  + C.C. \right) + \frac{\hbar}{4 \varpi}\dot h_{jk}(t')  \dot h_{jl}(t') \left(a_{k}a_{l}+a_{k}a_{l}^\dagger \right. \nonumber\\&& \left. +a_{k}^\dagger a_{l}
	+a_{k}^\dagger a_{l}^\dagger \right).
	\label{Q}
\end{eqnarray}
Expanding out the above interaction term for $j,k = 1,2$, we obtain the transition amplitude between the ground state $|0,0\rangle$ and the perturbed second excited states to be
\begin{eqnarray}
	C_{0\rightarrow 2^{(0)}} &=& - \frac{i \varpi}{\hbar\sqrt{2}} \int_{-\infty}^{+\infty} dt \, e^{2i\varpi t} \hbar \left( \dot h_{11}^2+ \dot h_{12}^2 \right)  \nonumber\\
	C_{0\rightarrow 2^{(1)}} &=& - \frac{i}{\hbar} \int_{-\infty}^{+\infty} dt \, e^{2i \varpi(1 + \Lambda)t} \hbar \left[ i  A(\Lambda) \dot h_{11}(t)
	\right. \nonumber\\&& \left. - B(\Lambda) \dot h_{12}(t)\right].\nonumber\\
	\rm  \nonumber\\
	C_{0\rightarrow 2^{(2)}} &=& - \frac{i}{\hbar} \int_{-\infty}^{+\infty} dt \, e^{2 i \varpi(1 - \Lambda)t}  \hbar \left[ i  C(\Lambda) \dot h_{11}(t) \right. \nonumber\\&& \left. - D(\Lambda) \dot h_{12}(t)\right]
	\label{trans_amp_02a}
\end{eqnarray}
where 
\begin{eqnarray}
A(\Lambda)=\frac{1}{\sqrt{2}} \left(1+ \Lambda \right), \quad B(\Lambda)=\frac{1}{\sqrt{2}} \left(\sqrt{\frac{3}{2}} \Lambda  + 1 \right), \nonumber\\ \quad	C(\Lambda)=\frac{1}{\sqrt{2}} \left(1 - \Lambda \right), \quad D(\Lambda)=\frac{1}{\sqrt2} \left(\sqrt{\frac{3}{2}} \Lambda - 1 \right).
\label{cons}
\end{eqnarray} 
Eq.(\ref{trans_amp_02a}) is the main working formula in this paper. With this general formula (\ref{trans_amp_02a}), we shall compute the corresponding transition probabilities from the relation
\begin{eqnarray}
	P_{0\rightarrow 2} =  |C_{0\rightarrow 2}|^{2}.
	\label{trans_prob}
\end{eqnarray}
At this stage, we would like to draw the attention to some interesting features about the second order terms in $h$  on the results obtained in \cite{sb}. The above transition amplitudes (\ref{trans_amp_02a}) show that $C_{0\rightarrow 2^{(1)}}$ and $C_{0\rightarrow 2^{(2)}}$ do not contain contribution from the second order term in $h$. The second order contribution in $h$ generates an additional transition between the ground state and the perturbed second excited state $\psi_2^{(0)}$, without altering the results for the transitions between the ground state and $	\psi_2^{(1)}$ or $	\psi_2^{(2)}$.
\section{Transition probabilities for different types of gravitational waves}
\noindent In this section we calculate the transition probabilities for different templates of gravitational wave-forms that are likely to be generated in runaway astronomical events. 
\noindent First we discuss the simplest scenario of periodic GW with linear polarization. This has the form
\begin{equation}
	h_{jk} \left(t\right) = 2f_{0} \cos{\Omega t} \left(\varepsilon_{\times}\sigma^1_{jk} + \varepsilon_{+}\sigma^3_{jk}\right)
	\label{lin_pol}
\end{equation}
where the amplitude varies sinusoidally with a single frequency $\Omega$. The transition probabilities in this case turn out to be
\begin{eqnarray}
P_{0\rightarrow 2^{(0)}} &=& 32 \pi^2 \varpi^2 f_{0}^4 \Omega^4 \left(\varepsilon_{+}{}^{2} +\varepsilon_{\times}{}^{2} \right)^2 [\delta \left(2 \varpi-2 \Omega\right) ]^2\nonumber\\ 
P_{0\rightarrow 2^{(1)}} &=&  \left( \pi f_{0} \Omega \right)^{2}  \left[A(\Lambda)^2 \varepsilon_{+}{}^{2}  + B(\Lambda)^2 \varepsilon_{\times}{}^{2} \right] \nonumber\\&&
\left[\delta \left(2  \varpi_{+} - \Omega \right) \right]^{2} \nonumber\\
P_{0\rightarrow 2^{(2)}} &=& \left( \pi f_{0} \Omega \right)^{2}  \left[C(\Lambda)^2 \varepsilon_{+}^{2} + D(\Lambda)^2 \varepsilon_{\times}^{2} \right] \nonumber\\&&
\left[\delta \left(2 \varpi_{-} - \Omega \right) \right]^{2}
\label{trans_probL}
\end{eqnarray}
\noindent where $\varpi_{+}=\varpi (1 + \Lambda)$ and $\varpi_{-} = \varpi (1- \Lambda)$. The restriction on the physical range of frequency $\left(0< \varpi < \infty \right) $ is imposed to drop the delta functions $\delta \left(2 \varpi_{+} + \Omega \right)$ and $\delta \left(2 \varpi_{-} + \Omega \right)$ that would appear in eq.(\ref{trans_probL}). The transition rates therefore take the form 
\begin{eqnarray}
   	\lim\limits_{T \rightarrow \infty} \frac{1}{T}P_{0\rightarrow 2^{(1)}} &= & 32 \pi^2 \varpi^2 f_{0}^4 \Omega^4 \left(\varepsilon_{+}{}^{2} +\varepsilon_{\times}{}^{2} \right)^2 \times \delta \left(2 \varpi-2 \Omega\right) \nonumber\\
	\lim\limits_{T \rightarrow \infty} \frac{1}{T}P_{0\rightarrow 2^{(1)}} &= & \left( \pi f_{0} \Omega \right)^{2}  \left[A(\Lambda)^2 \varepsilon_{+}{}^{2}  + B(\Lambda)^2 \varepsilon_{\times}{}^{2} \right]\nonumber\\&& \times \delta \left(2 \varpi_{+} - \Omega \right) 
\label{tm}\nonumber \\
	\lim\limits_{T \rightarrow \infty} \frac{1}{T}P_{0\rightarrow 2^{(2)}} &=&  \left( \pi f_{0} \Omega \right)^{2}  \left[C(\Lambda)^2 \varepsilon_{+}{}^{2}  + D(\Lambda)^2 \varepsilon_{\times}{}^{2}   \right] \nonumber\\&& \times \delta \left(2 \varpi_{-} - \Omega \right)
\label{trans_rate}
\end{eqnarray}
where we have used the relation 
\begin{eqnarray}
\delta(\varpi)=\left[ \int_{-\frac{T}{2}}^{\frac{T}{2}} dt \, e^{i \varpi t}\right] = T.
\label{time period}
\end{eqnarray}

\noindent Now looking at the expressions for  $A, B, C, D$ and the transition probabilities in eq.(s) (\ref{trans_probL}) and (\ref{trans_rate}), it is easy to see that the transition rates will be peaked around the frequencies $\Omega= 2 \varpi_{+}$ and $\Omega= 2 \varpi_{-}$ with an unequal strength. Further, transition probabilities induced by both the $+$ and $\times$ polarizations of the GW are affected by spatial noncommutativity. In other words, the orientation of the GW source and the detector do not play a crucial role any more to detect the spatial NC effect. Besides these two resonant points already observed in \cite{sb}, there is another resonant point at $\Omega=\varpi$ which arises due to the second order term in $h$. This was absent in \cite{sb}. This is a crucial result in this paper. The second order term in $h$ gives rise to the transition probability $P_{0\rightarrow 2^{(0)}}$. This is a  purely gravity induced effect. It is also obvious from the expressions of $A, B, C, D$ in eq.(\ref{cons}) that both linear and quadratic terms in the dimensionless NC parameter $\Lambda$ will appear in the transition probabilities (\ref{trans_probL}).

\noindent The characteristic NC parameter $\Lambda$, was estimated in \cite{cqg} where the stringent upper-bound $|\theta| \approx 4 \times 10^{-40} {\rm m}^{2}$ \cite{carol} for spatial noncommutativity was used. For reference mass and frequency, the values appropriate for fundamental phonon modes of a bar detector \cite{ncgw_4} can be used. This gives 
\begin{eqnarray}    \Lambda = \frac{m \varpi \theta}{2 \hbar} = 1.888 \left( \frac{m}{10^{3}{\rm kg}}\right) \left( \frac{\omega}{1{\rm kHz}}\right)     \label{dim_less_NC1} .    \end{eqnarray}
This is an interesting result since the estimated size of the characterestic NC parameter turns out to be of the order of unity in case of resonant bar detectors. This in turns gives the estimate for the characteristic NC frequency to be in the KHz range.


\noindent From the entire discussion so far we find a very interesting feature that both the $+$ and $\times$ polarizations includes the effects of the NC structure of space. Therefore the linearly polarized GW from a binary system with its orbital plane lying parallel or perpendicular to our line of sight can also be an effective test of the noncommutative structure of space. 

\noindent With these observations in place we now move on to compute the transition probabilities for circularly polarized GW. To proceed we take the simplest form of a periodic GW signal with circular polarization, which can be conveniently expressed as
\begin{equation}
	h_{jk} \left( t \right) = 2f_{0} \left[\varepsilon_{\times} \left( t \right) \sigma^1_{jk} + \varepsilon_{+}\left( t \right) \sigma^3_{jk}\right] 
	\label{cir_pol}
\end{equation}
with $\varepsilon_{+} \left( t \right)  = \cos \Omega t $ and $\varepsilon_{\times} \left( t \right)  = \sin \Omega t $ and $\Omega$ is the frequency of the GW. 
\noindent The transition rates in this case become
\begin{eqnarray}
\lim\limits_{T \rightarrow \infty} \frac{1}{T} P_{0\rightarrow 2^{(0)}} &=& 32 ~\pi^2~ \varpi^2 ~f_{0}^4 ~\Omega^2 ~~\delta(2\varpi)=0 \nonumber\\
\lim\limits_{T \rightarrow \infty} \frac{1}{T} P_{0\rightarrow 2^{(1)}} &=& \left ( \frac{ f_{0} \Omega}{\hbar} \right)^{2} \left [A(\Lambda)^2 +  B(\Lambda)^2  \right ] \delta \big(2 \varpi_{+} - \Omega \big)
\nonumber\\
\lim\limits_{T \rightarrow \infty} \frac{1}{T} P_{0\rightarrow 2^{(2)}} &=& \left ( \frac{ f_{0} \Omega}{\hbar} \right)^{2}  \left [C(\Lambda)^2 +  D(\Lambda)^2 \right ] \delta \big(2 \varpi_{-} - \Omega \big).\nonumber\\
\label{trans_prob_cir_pol1}
\end{eqnarray}
From the above results, we observe that the characteristics of the transition rates for linearly polarized GW holds for circularly polarized GW signals as well upto first order in $h$. Thus circularly polarized GW from a binary system can also serve as a deterministic probe for spatial noncommutativity. However if we consider the effect of the second order term in $h$, the transition probability $ P_{0\rightarrow 2^{(0)}}$ is absent for circularly polarized GW which is clearly different from that for the linearly polarized GW. This interesting result can be used to determine the type of polarization of the GW source. 



\noindent Now we proceed to investigate the status of our system interacting with aperiodic GW signals, which are basically generated from GW bursts. GW bursts can be modelled by taking a simple choice as 
\begin{eqnarray}
	h_{jk} \left(t\right) = 2f_{0} g \left( t \right) \left(\varepsilon_{\times}\sigma^1_{jk} + \varepsilon_{+}\sigma^3_{jk}\right)
	\label{lin_pol_burst}
\end{eqnarray}
containing both components of linear polarization. We further take a Gaussian form for the function $g(t)$

\begin{equation}
	g \left(t\right) = e^{- t^{2}/ \tau_{g}^{2}}
	\label{burst_waveform_Gaussian}
\end{equation}
with $\tau_g \sim \frac{1}{f_{max}}$, where $f_{max}$ is the maximum value of a broad range continuum spectrum of frequency. 
Note that at $t=0$, $g \left( t \right)$ goes to unity.
Now the Fourier decomposed modes of the GW burst can be written as 
\begin{eqnarray}
	h_{jk} \left(t\right) = \frac{f_{0}}{\pi} \left(\varepsilon_{\times}\sigma^1_{jk} + \varepsilon_{+}\sigma^3_{jk}\right)  \int_{-\infty}^{+\infty} \tilde{g} \left( \Omega \right) e^{- i \Omega t}  d \Omega 
	\label{lin_pol_burst_Gaussian}
\end{eqnarray}
where $\tilde{g} \left( \Omega \right) = \sqrt{\pi} \tau_{g} e^{- \left( \frac{\Omega \tau_{g}}{ 2} \right)^{2}}$ is the amplitude of the Fourier mode at frequency $\Omega$.

\noindent The transition probabilities induced by a GW burst can now be computed and read \cite{sb}
\begin{eqnarray}
P_{0\rightarrow 2^{(0)}} &= & 2 \varpi^6 f_{0}^4 \tau_{g}^4 e^{-\tau_{g}^2 \varpi^2}\left[\varepsilon_{+}^2 + \varepsilon_{\times}^2\right]^2 \nonumber\\
P_{0\rightarrow 2^{(1)}}  &=& \left[ 4 \sqrt{\pi}  f_{0} \tau_{g} \varpi_{+} \right]^{2} e^{- 2 \tau_{g}^2 \varpi_{+}^2}\nonumber\\&&\times \left( A(\Lambda)^2 \varepsilon_{+}{}^{2}  + B(\Lambda)^{2} \varepsilon_{\times}{}^{2}  \right)\nonumber
\label{trans_prob_Gaussian_burst}\\
P_{0\rightarrow 2^{(2)}} &= &\left[ 4  \sqrt{\pi}  f_{0} \tau_{g} \varpi_{-}\right]^{2} e^{-2 \tau_{g}^2 \varpi_{-}^2 } \nonumber\\&& \times\left( C(\Lambda)^2 \varepsilon_{+}{}^{2}  + D(\Lambda)^{2} \varepsilon_{\times}{}^{2} \right).
\nonumber\\
\label{trans_prob_Gaussian_burst1}
\end{eqnarray}

\noindent Lastly, we consider a modulated Gaussian function $g(t)$ of the form
\begin{equation}
	g \left(t\right) = e^{- t^{2}/ \tau_{g}^{2}}  \,  \sin \Omega_{0}t
	\label{burst_waveform_sine_Gaussian}
\end{equation}
which represents a more realistic model of the GW burst signal. The Fourier transform this function reads
\begin{eqnarray}
	\tilde{g} \left( \Omega \right) &=& 2 \pi \int_{-\infty}^{+\infty} g(t) e^{i \Omega t} d \Omega \nonumber\\ &=& \frac{i \sqrt{\pi} \tau_{g}}{2} \left[ e^{- \left(\Omega - \Omega_{0}\right)^{2}\tau_{g}^{2}/4} - e^{- \left(\Omega + \Omega_{0}\right)^{2}\tau_{g}^{2}/4} \right].\nonumber
\end{eqnarray}
From this waveform,
we get the transition amplitudes to be 
\begin{eqnarray}
	C_{0\rightarrow 2^{(1)}}&= & \frac{i}{2\sqrt{2}}~ \varpi^3 f_{0}^2 \tau_g^2\left( \varepsilon_{\times}^2+ \varepsilon_{+}^2\right)\nonumber\\&& \left[e^{-\frac{(\varpi-\Omega_0)^2\tau_g^2}{4}}- e^{-\frac{(\varpi+\Omega_0)^2\tau_g^2}{4}}\right]^2\nonumber\\
	C_{0\rightarrow 2^{(1)}}&= & \left[ e^{- \left(2 \varpi_{+} - \Omega_{0}\right)^{2}\tau_{g}^{2}/4} - e^{- \left(2 \varpi_{+} + \Omega_{0}\right)^{2}\tau_{g}^{2}/4} \right]\nonumber\\&&\times  \left( 2 \sqrt{\pi}  f_{0} \varpi_{+} \tau_{g} \right) \left(A(\Lambda) \varepsilon_{+}  + B(\Lambda) \varepsilon_{\times} \right)\nonumber
	\label{gwptt1}
\end{eqnarray}
\begin{eqnarray}
	C_{0\rightarrow 2^{(2)}}&= & \left[ e^{- \left(2 \varpi_{+} - \Omega_{0}\right)^{2}\tau_{g}^{2}/4} - e^{- \left(2 \varpi_{+} + \Omega_{0}\right)^{2}\tau_{g}^{2}/4} \right] \nonumber\\&&\times  \left( 2 \sqrt{\pi}  f_{0} \varpi_{+} \tau_{g} \right) \left(C(\Lambda) \varepsilon_{+}  + D(\Lambda) \varepsilon_{\times} \right).\nonumber\\
	\label{gwptt2}
\end{eqnarray}
 The corresponding transition probabilities are
\begin{eqnarray}
	P_{0\rightarrow 2^{(0)}} & = & \frac{1}{8} \varpi^6 f_0^4 \tau_g^4 \left( \varepsilon_{\times}^2+ \varepsilon_{+}^2\right)^2\nonumber\\&&
	\left[e^{-\frac{(\varpi-\Omega_0)^2\tau_g^2}{4}}- e^{-\frac{(\varpi+\Omega_0)^2\tau_g^2}{4}}\right]^4\nonumber\\
	P_{0\rightarrow 2^{(1)}} & = & \left[ e^{- \left(2 \varpi_{+} - \Omega_{0}\right)^{2}\tau_{g}^{2}/4} - e^{- \left(2 \varpi_{+} + \Omega_{0}\right)^{2}\tau_{g}^{2}/4} \right]^{2} \nonumber\\&&\times  \left( 2 \sqrt{\pi}  f_{0} \varpi_{+} \tau_{g} \right)^{2} \left(A(\Lambda)^2 \varepsilon_{+}{}^{2}  + B(\Lambda)^{2} \varepsilon_{\times}{}^{2} \right) 
	\label{trans_prob_sine_Gaussian_burst}\nonumber\\
	P_{0\rightarrow 2^{(2)}} & = & \left[ e^{- \left(2 \varpi_{-} - \Omega_{0}\right)^{2}\tau_{g}^{2}/4} - e^{- \left(2 \varpi_{-} + \Omega_{0}\right)^{2}\tau_{g}^{2}/4} \right]^{2}\nonumber\\&& \times  \left( 2 \sqrt{\pi}  f_{0} \varpi_{-} \tau_{g} \right)^{2} \left(C(\Lambda)^2 \varepsilon_{+}{}^{2}  + D(\Lambda)^{2} \varepsilon_{\times}{}^{2} \right).\nonumber\\
	\label{trans_prob_sine_Gaussian_burst1}
\end{eqnarray}
At low operating frequency of the detector, 
 the two exponential terms in the transition amplitudes are almost equal and hence cancel. Therefore, the transition probabilities are reduced considerably. 
The other extreme is when $2 \varpi_{+} - \Omega_{0} = \Delta \varpi_{+}$ and  $2 \varpi_{-} - \Omega_{0} = \Delta \varpi_{-}$ with $\frac{\Delta \varpi_{+}}{\varpi_{+}} << 1$ and $\frac{\Delta \varpi_{-}}{\varpi_{-}} << 1$ respectively. This yields \cite{sb}
\begin{eqnarray}
    P_{0\rightarrow 2^{(1)}}  \approx  \frac{1}{8} \varpi^6 f_0^4 \tau_g^4 \left( \varepsilon_{\times}^2+ \varepsilon_{+}^2\right)^2 e^{-16 \varpi^2 \tau_g^2}\nonumber\\
	P_{0\rightarrow 2^{(1)}}  \approx   e^{- \left(\Delta \varpi_{+} \right)^{2}\tau_{g}^{2}/2}  \left(2 \sqrt{\pi}  f_{0} \varpi_{+} \tau_{g} \right)^{2}\nonumber\\ \left(A(\Lambda)^2 \varepsilon_{+}{}^{2}  + B(\Lambda)^{2} \varepsilon_{\times}{}^{2} \right)
	\nonumber\\ 
	P_{0\rightarrow 2^{(2)}}  \approx   e^{- \left(\Delta \varpi_{-} \right)^{2}\tau_{g}^{2}/2}  \left(2  \sqrt{\pi}  f_{0} \varpi_{-} \tau_{g} \right)^{2} \nonumber\\\left( C(\Lambda)^2\varepsilon_{+}{}^{2}  + D(\Lambda)^{2} \varepsilon_{\times}{}^{2} \right).
	\label{trans_prob_sine_Gaussian_burst_2}
\end{eqnarray} 

\section{Conclusion}
\noindent In this paper, we have extended our earlier calculations of the probabilities of transitions between the energy levels of a harmonic oscillator induced by gravitational waves in a spatial noncommutative framework including terms upto second order in $h$. We find that apart from the usual transition \cite{sb}, there is an additional transition between the ground state and the perturbed second excited states. The probabilities of the other transitions that were present earlier \cite{sb} remained unaffected by the inclusion of terms second order in $h$.   An interesting result that we observe in our present investigation is that the additional transition probability observed in case of linearly polarized gravitational wave (due to the inclusion of second order terms in $h$ in the Hamiltonian) is absent in case of circularly polarized gravitational wave. This result can in principle be used to determine the type of polarization of the gravitational wave source.



\end{document}